# Evolution of SLiM-mediated hijack functions in intrinsically disordered viral proteins

Juliana Glavina[1,2], Nicolas Palopoli[3,4] and Lucía Beatriz Chemes[1,2,5,*]

Affiliations:

[1] Instituto de Investigaciones Biotecnológicas, Universidad Nacional de San Martín (UNSAM) – Consejo Nacional de Investigaciones Científicas y Técnicas (CONICET).

[2] Escuela de Bio y Nanotecnologías (EByN), Universidad Nacional de San Martín, Buenos Aires, Argentina.

[3] Departamento de Ciencia y Tecnología, Universidad Nacional de Quilmes.

[4] Consejo Nacional de Investigaciones Científicas y Técnicas (CONICET).

[5] Structural and Computational Biology Unit, European Molecular Biology Laboratory, Heidelberg, Germany.

[*] to whom correspondence should be addressed: lchemes@iib.unsam.edu.ar

# Summary Bullet Points

- Intrinsically disordered proteins (IDPs) play a critical role in viral adaptation to new environments

- Viral short linear motifs (SLiMS) evolve fast and convergently. Functionally-coupled SLiMs co-evolve, indicating that viral SLiMs are subject to functional selection.

- Changes in viral SLiMs are associated with phenotypic traits and host switch events, highlighting that SLiMs are likely to be key players in viral adaptive evolution.

- The sequence of viral SLiMs may be fine-tuned for high binding affinities that allow outcompeting host interactions.

- Tethering of multiple SLiMs by disordered linkers is an alternative strategy for viral hijack. Changes in SLiMs within the linkers may enable the rewiring of multivalent complexes necessary for viral adaptation.

# Abstract

Viruses and their hosts are involved in an "arms race" where they continually evolve mechanisms to overcome each other. It has long been proposed that intrinsic disorder provides a substrate for the evolution of viral hijack functions and that short linear motifs (SLiMs) are important players in this process. Here, we review evidence in support of this tenet from two model systems: the papillomavirus E7 protein and the adenovirus E1A protein. Phylogenetic reconstructions reveal that SLiMs appear and disappear multiple times across evolution, providing evidence of convergent evolution within individual viral phylogenies. Multiple functionally related SLiMs show strong co-evolution signals that persist across long distances in the primary sequence and occur in unrelated viral proteins. Moreover, changes in SLiMs are associated with changes in phenotypic traits such as host range and tropism. Tracking viral evolutionary events reveals that host switch events are associated with the loss of several SLiMs, suggesting that SLiMs are under functional selection and that changes in SLiMs support viral adaptation. Fine-tuning of viral SLiM sequences can improve affinity, allowing them to outcompete host counterparts. However, viral SLiMs are not always competitive by themselves, and tethering of two suboptimal SLiMs by a disordered linker may instead enable viral hijack. Coevolution between the SLiMs and the linker indicates that the evolution of disordered regions may be more constrained than previously thought. In summary, experimental and computational studies support a role for SLiMs and intrinsic disorder in viral hijack functions and in viral adaptive evolution.

# Introduction

***Functional modules within intrinsically disordered regions***. Intrinsically disordered proteins (IDPs) and protein regions (IDRs) make up over one third of eukaryotic proteomes [1] and are enriched within many viral families [2]. IDPs lack a consolidated three-dimensional structure, and instead exist as heterogeneous conformational ensembles with different degrees of structure and compaction. The relevant functional roles of IDPs have been recognized over the last couple of decades. A major functional module found within IDPs and IDRs are short linear motifs or SLiMs [3–5]. SLiMs are short sequence elements between 3 and 10 amino acids in length that usually present a few highly conserved amino acids that dictate their binding specificity. SLiMs are located in intrinsically disordered regions, where they are readily accessible for interactions with other proteins. SLiMs play key roles in signaling by mediating protein-protein interactions (ligand binding function) and regulating protein stability (degron function), cellular localization (targeting function) or post-translational modifications (e.g. phosphorylation, glycosylation). Reversible post-translational modifications of SLiMs can act as signaling switches, and multivalent binding mediated by SLiMs can cooperate in the assembly of signaling platforms. Another main function of IDRs is as flexible tethers that join together ordered domains and/or disordered SLiMs [6,7]. Tethering is essential for many cellular functions by allowing the coupling of protein modules [8–10].

***Intrinsic disorder and the evolution of viral hijack functions.*** Viruses are obligate intracellular parasites that need to subvert host cell regulation in order to complete their life cycle. Viruses and their hosts are involved in a sustained "arms race" where they evolve mechanisms to overcome each other [11–13]. As a result, viruses manage to evade the host antiviral response [12,14] and gain the ability to infect new hosts by crossing species barriers [15]. Intrinsic disorder has been proposed to play a key role in viral hijack mechanisms [16–18] and is thought to facilitate viral evolution and adaptation [19]. Multiple hypotheses have been put forward to explain how intrinsic disorder facilitates viral subversion of the host cell. For example, structural heterogeneity is proposed to allow IDPs and IDRs to evolve at a faster rate, which may promote viral adaptation by allowing the acquisition of novel traits with minimal deleterious impact [20,21]. A second proposal is that protein disorder allows encoding of multiple functions in a short stretch of sequence using SLiMs, helping overcome the small size and limited coding capacity of viral genomes. The above mentioned hypotheses, that have been referred to as the "messiness" provided by intrinsic disorder [22] are considered to be key for explaining viral innovation. However, few of these hypotheses have been comprehensively tested.

***SLiMs as key players in viral hijack***. Viruses subvert host cell function by mimicking host components, with SLiM mimicry being a main strategy of viral hijack [23]. The evolutionary plasticity of SLiMs places them as key candidates for being mediators of viral adaptive evolution (Box 1) [24,25], and recent large-scale studies confirm the widespread use of SLiM mimicry by viruses [26–29]. Effective viral hijack could depend on the evolution of higher affinity SLiMs, on the establishment of cooperative multivalent interactions mediated by SLiMs and/or domains, or both.

***Viral oncoproteins from DNA tumor viruses***. The papillomavirus E7 protein and the adenovirus E1A protein are considered viral oncoproteins due to their ability to induce tumor formation and transform mammalian cells [30,31]. E7 and E1A are intrinsically disordered and densely packed with SLiMs that mediate interactions with multiple binding partners (for comprehensive reviews, see [32] and [33]) (Figure 1A,B). E7 is a modular protein composed of a disordered N-terminal domain (E7N) and a globular C-terminal domain (E7C) (Figure 1A) [34–38]. E1A is almost entirely disordered except for a small folded domain at its C-terminus [39] (Figure 1B). E7 and E1A hijack the cell cycle by binding to the retinoblastoma (Rb) protein using two SLiMs, the LxCxE and E2F motifs (Figure 1 A,B). This interaction inactivates Rb and leads to activation of E2F transcription factors [40]. The E7 and E1A proteins are disordered as determined through nuclear magnetic resonance, circular dichroism and other biophysical methods [36–38,41–43]. This, together with the wealth of sequence and phenotypic information available for both viral families, makes them excellent model systems to study the role of disorder and SLiMs in viral hijack.

Here, we review how studies using the E1A and E7 model systems helped elucidate the mechanisms through which disorder and SLiMs facilitate the evolution of viral hijack functions. In the first two sections, we show how the combined use of sequence analysis and reconstruction of SLiM evolutionary history helped uncover patterns of SLiM evolution. Correlating these patterns with phenotypic information revealed associations between SLiMs and viral phenotypes. We further address the molecular mechanisms of viral mimicry, and show how fine-tuning of SLiM sequences can lead to optimization of affinity in viral motifs when compared to host SLiMs. Finally, we extend the analysis to multivalent interactions involving the tethering of SLiMs by flexible linkers, and provide an outlook of how these results integrate with results from other viral systems.

# Main Text

## Evolution of SLiMs in viral intrinsically disordered proteins.

SLiMs are able to evolve quickly, since a single amino acid substitution can lead to their *de novo* appearance or disappearance in a protein sequence [5,25]. The *de novo* appearance of SLiMs is also termed convergent evolution (Box 1). SLiMs can function cooperatively, and functionally coupled SLiMs, such as for example SLiMs which mediate binding to the same protein partner, were proposed to coevolve [44]. While more is known about host SLiMs [3,45,46], few studies have addressed how viral SLiMs change across evolution, the extent to which they evolve convergently or whether their evolutionary patterns are coupled. The presence of multiple SLiMs and the wealth of E7 and E1A protein sequences makes them good model systems to draw correlations between the evolutionary patterns of SLiMs across viral phylogenies and viral phenotypes.

Studies performed in E7 and E1A revealed that the abundance of SLiMs in both proteins varies broadly, ranging from SLiMs present in all sequences to SLiMs present in just a few [47,48]. SLiMs are distributed unevenly across viral types, suggesting that their evolution follows specific patterns, which could be related to adaptations occurring in specific viral types [47,48].

The evolutionary history of SLiMs can be traced across the viral phylogeny by scoring their presence and absence in ancestral protein sequences reconstructed from observed sequences. The gain of a SLiM is defined by a change from absence in the ancestral sequence to presence in the descendant sequence, and the opposite defines a SLiM loss (Figure 2A). The evolutionary history of SLiMs in the E7 [47] and E1A proteins [49] revealed multiple independent gains and losses in recent and deep branches of the viral tree (Figure 2A). These observations provide direct evidence that convergent evolution and SLiM loss occur within individual phylogenies. The frequency of these gain and loss events suggested that the viral SLiMs evolve quickly [47]. A calculation based on the E1A protein confirmed this hypothesis, showing that the rate of change in SLiM-mediated protein-protein interactions is ~1 interaction per million years [49], a value four to five orders of magnitude faster than in globular proteins, where the higher bound is estimated to be ~$1.10^{-5}$ interactions per million years [50].

An important question is whether SLiMs evolve independently, or whether their evolution is coupled. One could expect that SLiMs with shared functions showed similar evolutionary patterns, for example in multivalent interactions where several SLiMs mediate binding to the same protein partner. To address this question, it is necessary to quantify

SLiM coevolution. At the amino acid level sequence coevolution can be calculated by quantifying correlated changes between two positions belonging to a SLiM in a multiple sequence alignment (Figure 2B) [51]. Associations between pairs of SLiMs can also be quantified by scoring the probability that their co-occurrence or their coordinated gain/loss occurs by chance. The E7 and E1A proteins show sequence coevolution between amino acids located within SLiMs [48,52]. The E7 and E1A proteins also show multiple co-occurrences and correlated gain/loss events between pairs of SLiMs [47,49] (Figure 2C). Most importantly, many of these correlations were found for SLiMs that mediate binding to the same protein partner. For example, the IDMBR and TRAM-CBP motifs of E1A, which cooperate to mediate binding to the CREB binding protein (CBP) [41,53–55], showed a high frequency of co-occurrences [49] (Figure 2D). Also, four SLiMs present in E7 and E1A that mediate binding to Rb (E2F, LxCxE, Acidic region and CKII phosphorylation site) [54,56–59] show multiple co-occurrences in both proteins and the gains and losses of these SLiMs are correlated in the E1A protein [47,49] (Figure 2D). These associations are robust, not only due to the correlated SLiM gain/loss events mentioned above, but also given that coevolution signals persist even when the spacing between these SLiMs in the E1A and E7 proteins changes from 10 to 75 amino acids [48,49,52]. Taken together, these studies provide evidence that viral SLiMs with shared functions evolve coordinately.

**Importance of SLiMs in viral adaptive evolution**

Many of the molecular interactions between the virus and the host are mediated by SLiMs. One important question is whether changes in SLiMs facilitate viral adaptation to different environments by leading to changes in phenotype, a process also called adaptive evolution [19] (Box 1). While the presence of specific SLiMs has been associated with viral phenotypes (for a review see [24]), and some studies provided evidence that SLiMs can be under positive selection (Box 1), a process linked to adaptive evolution [60,61], comprehensive studies tracking how changes in SLiMs correlate with changes in phenotype are rare.

Similar to the analysis of SLiM co-occurrences, the association between SLiMs and viral phenotypes such as tissue or host tropism (Box 1) can be inferred from statistical associations between the presence of a SLiM and a given viral phenotype. Papillomaviruses have different tissue tropism, causing specific lesions such as cutaneous papillomas, mucosal papillomas or fibropapillomas. In E7, the absence or presence of specific SLiMs is associated with tissue tropism [47] (Figure 3A). Papillomaviruses also infect a wide range of vertebrate hosts [62], and the presence and absence of different sets of SLiMs correlates with host tropism [47] (Figure 3A). Moreover, in papillomaviruses the presence of a high affinity LxCxE motif in the E7 protein correlates with a higher risk for the development of cervical cancer [58,63]. Altogether, this evidence suggests that E7 SLiMs play a role in viral adaptive evolution by determining papillomavirus pathogenicity and viral tropism.

The role of SLiMs in viral adaptive evolution can also be inferred by associating changes in the viral protein SLiMs with viral evolutionary events (Box 1). While adenovirus evolution is largely driven by cospeciation, other evolutionary events including duplication, host switch and partial extinction also shape the evolutionary history of adenoviruses [49] (Figure 3B). These results agree with case studies [64,65] and with a quantitative study that inferred cospeciation and host switch events between adenoviruses infecting simians and humans in Africa [15]. The correlation between viral evolution and changes in SLiMs reveals that SLiM gain/loss events are associated with adenovirus evolutionary events [49]. Host switch is associated with the gain of one and the loss of four SLiMs (Figure 3C) and partial extinction is associated with the loss of one SLiM (Figure 3C). While these studies do not allow to determine whether changes in SLiMs follow or precede the evolutionary events, the results support the conclusion that changes in E1A SLiMs contribute to viral adaptive evolution.

**SLiM features that enhance viral hijack**

Host cell hijack by E7 or E1A relies on outcompeting cellular interactions with pocket proteins (Rb, p107 and p130). This is achieved through viral variants of the LxCxE motif that mimic the host SLiMs but have optimized affinity allowing high affinity binding to Rb [58,66–68]. The E7 LxCxE motif binds to Rb with nanomolar affinity, higher but still comparable with LxCxE variants in other viral proteins like E1A or simian virus large T antigens. In contrast, LxCxE motifs from host proteins tend to bind Rb with weaker, micromolar, affinity [69] (Figure 4A). The LxCxE motif can promote viral hijack by providing a primary docking site that acts along with a secondary site to outcompete the E2F transcription factor and disrupt the Rb-E2F complex, or can act by directly competing with multiple host factors bound to the LxCxE cleft in Rb, such as HDAC, KDM5A and others [69].

Recent biophysical and structural studies helped uncover the molecular basis for fine-tuning of affinity and specificity of viral versus host LxCxE motifs, which occurs through changes in both the core motif and the flanking regions. The LxCxE motif binds into a shallow groove in Rb termed the LxCxE cleft [66]. Residues LYCYE (positions 22-26) constitute the core motif in E7. The conserved residues L22 and C24 bind to hydrophobic pockets of the LxCxE cleft, while E26 establishes hydrogen bonds to two main chain atoms of Rb [70] (Figure 4B). The variable positions of the core motif are important determinants of Rb-binding affinity [67]. Aromatic amino acids in these positions, such as the tyrosines found in the E7 motif and the phenylalanine found in the SV40 large T antigen and KDM5A motifs (Figure 4A), enhance binding affinity. This effect is likely due to stacking interactions (Figure 4B) that shield hydrogen bonds between the motif and the pocket protein from the solvent [69].

While the core LxCxE motif is essential for binding to Rb and other pocket proteins, the N- and C-terminal regions flanking the core motif play additional roles in fine-tuning the affinity of LxCxE-mediated interactions. For example, a hydrophobic residue located two residues after the core motif, i.e. in position (+2), binds to a broad hydrophobic pocket present in Rb and p107 [66,71]. Although the conservation of this residue in viral and cellular LxCxE proteins suggests an optimal position at (+2), the broad size of the binding pocket allows different hydrophobic residues located at (+2) and (+3) to bind in different orientations, providing fine-tuning of the binding affinity (for example, it is found at (+3) in adenovirus E1A and human ARID4A) [69,70] (Figure 4A). An acidic residue in position (-1) (Figure 4A) is conserved in E7 and other viral proteins. This extra negative charge may increase binding affinity [67], either by interactions within the motif or with Rb [69]. The Acidic region, a stretch of acidic residues C-terminal to the motif (Figure 4A), also enhances binding to Rb in viral proteins, and phosphorylation of serines within that region further enhances affinity; for example, phosphorylation of S31 and S32 in E7 and of S132 in E1A produce a five-fold increase in affinity for Rb [58,59,72]. Kinetic studies revealed that this affinity enhancement is achieved through stabilizing electrostatic interactions with positively charged residues surrounding the LxCxE binding cleft that speed up the association reaction [72]. Indeed, basic residues are depleted around viral LxCxE motifs compared with the host cell proteins (Figure 4A,C). Substitutions in core and flanking regions also determine the binding specificity for different pocket proteins. For example, the non-canonical LxSxE motif in the human LIN52 protein introduces a non-optimal serine with a more polar side chain and a second phosphorylation site at position (+6). These substitutions favor interactions with the p107/p130 proteins but prevent binding to Rb, which lacks the binding pocket for the phosphorylated residue [71]. In summary, structural, sequence and biophysical evidence indicates that the viral proteins fine-tune the LxCxE motif sequence at core and flanking positions in order to optimize binding to pocket proteins and efficiently compete with host factors.

**Long-range cooperativity and multivalent binding in viral IDPs**

The E7 protein provides an outstanding example of viral hijack where fine tuning of an individual SLiM (the LxCxE motif) confers optimized binding affinity over all host SLiMs. However, viral hijack is often mediated by multivalent protein interactions involving more than one SLiM, or combinations of SLiMs and domains [23,24]. E1A provides a model system where two SLiMs (E2F and LxCxE) tethered by a flexible linker are required for binding to Rb and displacement of E2F transcription factors [56,57] (Figure 2D). The enhancement provided by tethering in E1A is amongst the highest achieved in related protein-protein or protein-DNA interactions [73,74], suggesting that this function is optimized. While a family of E1A linkers displays large-scale variations in sequence composition, patterning, and length, compensatory changes in the linkers lead to conserved dimensions that support optimal tethering (Figure 5A). This evolutionary mechanism that conserves the linker dimensions while allowing large variations in the linker sequences is called "conformational buffering" [59].

Optimal tethering of the E2F and LxCxE SLiMs by a flexible linker is highly conserved across E1A proteins. The analysis of SLiMs and linkers from different E1A proteins revealed that the SLiMs and the linker show correlated changes, suggesting that they coevolve (Figure 5B). For example, in E1A proteins infecting bats, mutations that make the linker suboptimal are compensated by mutations that increase the binding affinity of the SLiMs, preserving high affinity binding of E1A to Rb (Figure 5B). This is an indication that the selection pressure to preserve binding to Rb is present for E1A proteins infecting bats [59]. On the contrary, E1A proteins infecting bovines harbor very short linkers and very low affinity SLiMs, which leads to the loss of the ability to displace E2F transcription factors bound to Rb (Figure 5B). This is an indication that the selection pressure to preserve binding to Rb is absent for E1A proteins infecting bovines, leading to the combined loss of optimal SLiMs and linker [59]. These examples provide evidence that SLiMs and linkers that play the same function (Rb binding) are under joint selection pressure, and that selection pressure can be present even in highly variable intrinsically disordered regions.

**General principles governing the evolution of viral hijack functions mediated by intrinsically disordered regions and SLiMs**

Many disordered regions from DNA and RNA viruses are rich in SLiMs that mediate viral hijack [23,26]. For example, IDRs in the Nef protein from HIV harbor a myristoylation motif that targets Nef to the membrane, an SH3 motif that mediates interactions with Src family kinases, and dileucine and diacidic motifs that mediate endosomal trafficking [23,75,76]. IDRs in the Epstein Barr Virus LMP1 protein contain multiple SLiMs that mediate the activation of NF-κB signaling, block apoptosis and drive cell proliferation [77], and IDRs of the paramyxovirus phosphoprotein contain a *soyuz* motif that mediates binding to the viral N protein [78], a STAT-1 binding motif that antagonizes interferon signaling [79,80] and is also used by poxviruses [81], and a nuclear export signal (NES) [80]. SLiM predictions over viral proteomes and case studies show that convergent evolution of SLiMs is widespread across DNA and RNA viruses [79,82–84], and that many pairs of SLiMs co-occur [26], suggesting that the SLiM coevolution uncovered in E7 and E1A may be general for many viral proteins.

*In vitro* evolution studies [85] (Box 1) support a role for IDRs and SLiMs in viral adaptation. In Dengue virus, mutations increasing fitness accumulate at IDRs and hotspot mutations in a glycosylation motif of the E protein correlate with adaptation to human versus mosquito cells [86]. For the foot and mouth disease virus (FMDV) VP1 protein, *in vitro* exposure to soluble integrin decoys leads to an enrichment of mutations in the VP1 RGD motif, while exposure to antibodies leads to the fixation of mutations in an antigenic loop proximal to the RGD motif [60,85]. Changes in SLiMs also correlate with viral phenotypes in natural infections. For example, changes in the PDZ and LxCxE SLiMs of the E6 and E7 proteins correlate with human papillomavirus oncogenicity [58,87]. In influenza virus, changes in the PDZ motif of the NS1 protein correlate with pathogenicity [88], while the number of glycosylation motifs of the hemagglutinin (HA) protein correlate with the immunogenicity of engineered HA proteins [89] and natural viral strains [90]. The presence of a PDZ motif is also associated with virulence of rabies strains [91], and multiple SLiMs predict response to therapy in HIV [92]. Positive selection, another proxy for adaptive evolution, was detected within IDRs of several proteins from herpesvirus [93], hepatitis E virus [94] and FMDV [85]. Therefore, evidence from many viral pathogens supports the hypothesis that IDRs and the SLiMs within them are key players in viral adaptive evolution and pathogenesis.

## Conclusions

Hosts and pathogens are involved in a sustained race to adapt and counter-adapt to each other [11]. This adaptation can be mediated by globular [12,95] and disordered [13,85,96] regions of host and viral proteins. In this review, we use two model systems to highlight how the use of intrinsic disorder and SLiMs by viruses [2] provides a selective advantage for viral adaptation. The high evolutionary rate of disordered regions [21,97] allows for rapid sequence changes to occur. Viral IDRs are subject to positive selection [93] and present high rates of variation [98,99], supporting a role for disordered regions in viral adaptation.

SLiM mimicry constitutes a major mechanism of viral interference with host function. SLiMs are highly evolvable elements present in host [46] and pathogen [24] proteomes. Many evolutionary patterns recognized in host SLiMs [25,45,46,100] are shared by pathogen SLiMs: SLiMs can appear de novo and evolve convergently across pathogen types [26] and within pathogen families [47,49]. Viral SLiMs evolve at a fast rate [49] and may undergo periods of accelerated evolution followed by fixation [60,85,90,94]. The systematic analyses performed on the E7 and E1A proteins reveal strong signatures of coevolution among SLiMs and between SLiMs and viral host tropism and highlight how functionally coupled SLiMs coevolve [47,49] (Figures 2 and 3). These signatures, together with large scale studies of SLiM co-occurrence [26] and relationships established in other viral families [24] provide evidence that SLiM co-evolution contributes to viral adaptation. Viral SLiM mimics target evolutionarily constrained SLiM -binding domains of host proteins [26,47,49,82,83], which may help pathogens to minimize the host's ability to counter-adapt. These advantages may be finely balanced by the need to avoid an antiviral response, since SLiMs can be recognized as epitopes by the host immune system [24].

The close inspection of host and viral SLiMs indicates that fine-tuning of binding affinity and specificity occurs at multiple levels [101] which include changes in core [67], wildcard [70,102] and flanking [72,103] residues (Figure 4) as well as conformational preferences such as secondary structure propensity [104,105]. Pathogen SLiMs may evolve higher affinities than host SLiMs by fine tuning the sequence of the SLiM and its flanking regions [58,69,70,106] in order to achieve specificity over host interactions, such as the enhancement of LxCxE affinity by tuning electrostatic interactions [72]. Electrostatic enhancement is present in other SLiM-mediated interactions [102,103] and the high proportion of charged residues present in IDRs [107] may facilitate this process. The increased affinity of the viral LxCxE SLiMs might be an evolutionary solution to provide the binding specificity necessary to outcompete host interactions and target pocket proteins early during the infection cycle, at very low expression levels. On the other hand, the limited affinity of cellular SLiMs might be an answer to the need for fast regulation of protein interactions [69,108].

Additional SLiM-mediated hijack mechanisms include increasing avidity through the creation of SLiM clusters [109] or enhancement of affinity through the establishment of multivalent interactions as seen in E1A [59] (Figure 5). The high affinity of E1A for Rb is likely required for E1A to subvert the host cell cycle at the low expression levels during the early stages of viral infection. The disordered E1A linker harbors the TRAM-CBP motif, which binds to CBP and mediates Rb acetylation and the MYND motif, which binds to the BS69 transcriptional co-repressor (Figure 5C). This allows E1A to act as a multivalent binding scaffold that mediates the formation of higher order complexes with transcriptional regulatory activity [41,54]. The E1A linker also mediates interactions with proteins that help suppress the host antiviral response [14]. Therefore, while the conserved tethering mechanism allows the linker to perform the core function of binding to Rb (Figure 5B), the variability observed in the E1A linker sequence may produce gains or losses of SLiMs that allow E1A proteins to rewire their interaction network, a process that may be required for adaptation when infecting a new host (Figure 5B) [49]. [49]

In the future, we envision that the development of tools that allow to reconstruct the evolutionary history of intrinsically disordered regions [46], where traditional alignment-based methods fail, will be closely integrated with better knowledge of the phylogenetic relationship of viral families and with *in-vitro* evolution assays to focus on improving our understanding of the mechanisms through which intrinsic disorder facilitates viral evolution. This knowledge may also lead to useful insights for pathogen control.

## Figures and Figure Legends

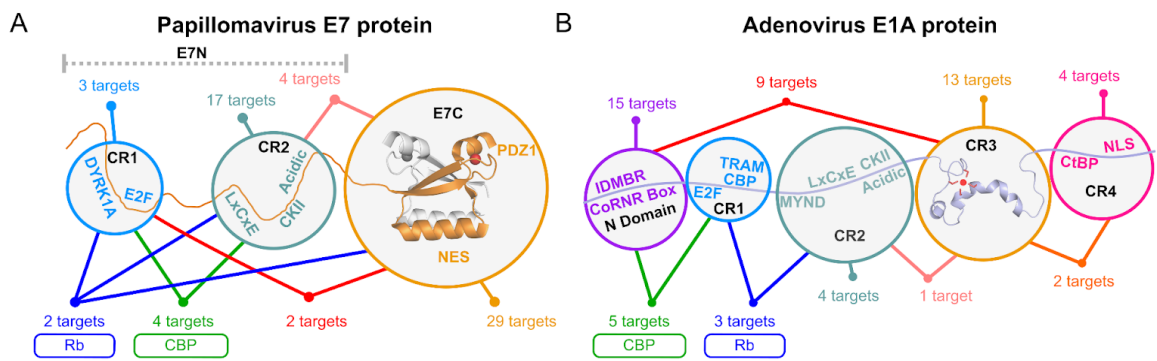

**Figure 1. Intrinsic disorder and SLiMs within papillomavirus E7 and adenovirus E1A proteins.** A-B) Schematic representations of E7 (A) and E1A (B) proteins. The approximate location of SLiMs and conserved regions (CR) and domains is indicated by labels and colored circles respectively. The size of each circle does not correlate with the size of the regions/domains. The disordered regions are indicated with a solid orange and light blue color line for the E7 and E1A proteins respectively. The E7 NMR structure of the HPV45 E7C domain (PDB 2F8B) and the modeled E1A CR3 domain are depicted as orange and light blue cartoons respectively and the coordinated zinc atom is shown as a red sphere for both proteins. The E7N domain comprises the CR1 and CR2 regions and is indicated with a dotted gray line. The number of targets mapped as binding to one or multiple domains are indicated for both proteins. Targets common for both proteins, CREB binding protein (CBP) and the retinoblastoma protein (Rb), are highlighted. For SLiM definitions and regular expressions of the E7 and E1A proteins see [47] and [48]. The E7 and E1A diagrams were adapted with permission from [52] and [48].

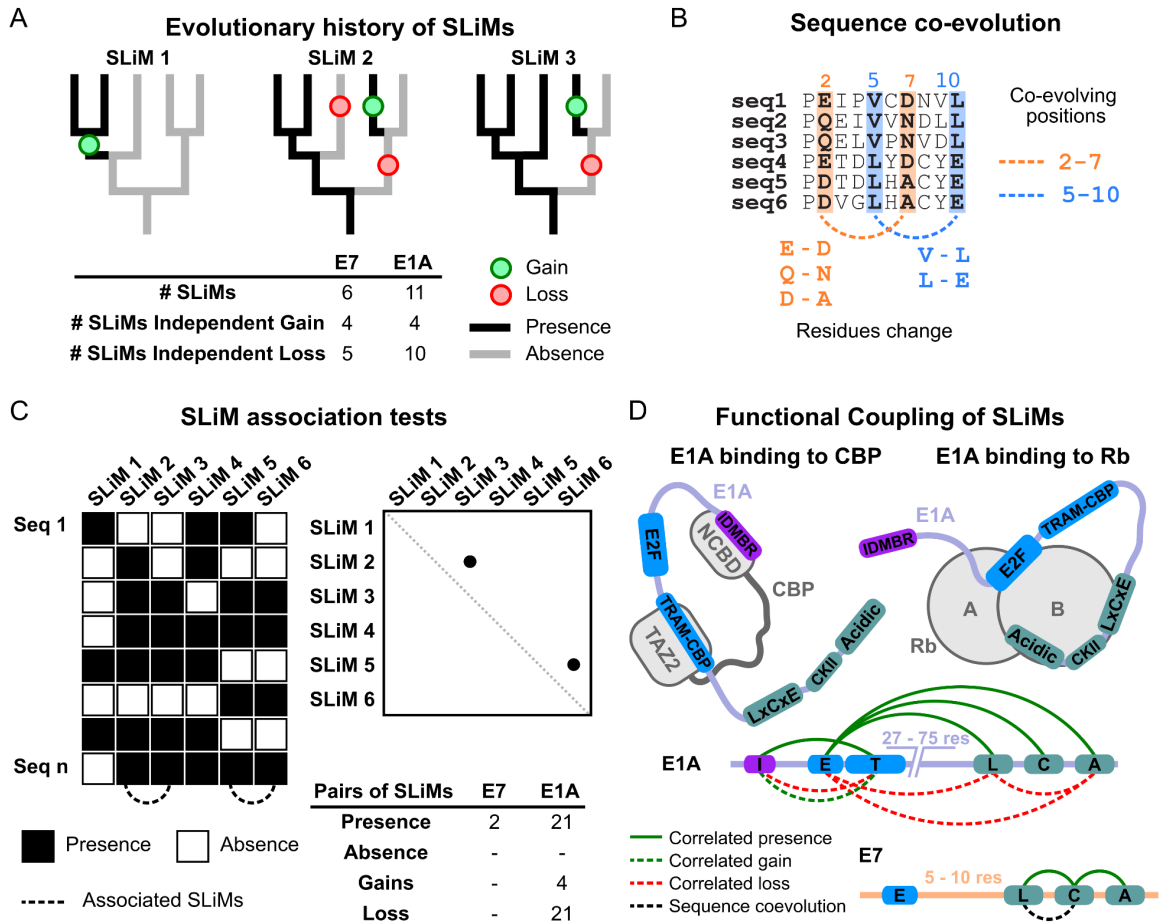

**Figure 2. Functional coupling of SLiMs can be inferred by sequence coevolution and correlated occurrences. A)** The evolutionary history of SLiMs can be traced on the viral phylogenetic tree, where the presence and absence of each SLiM is depicted as black and gray branches respectively. Gain and loss events are depicted by green and red circles respectively. A representative example for correlated evolutionary patterns is presented for SLiM2 and SLiM3. Results for E7 and E1A are summarized below the diagrams. **B)** Correlations between changes at two positions of a multiple sequence alignment can be used to infer the co-evolution of residue pairs. An example of correlated changes is shown for positions 5 and 10 (light blue) and 2 and 7 (orange). **C)** Functional coupling between a pair of SLiMs can be scored through a hypergeometric test, which assesses the probability that the observed pattern occurs by chance. In the figure, SLiMs 2 and 3, and SLiMs 5 and 6 show co-occurrence. The results can be depicted as a dot-plot, where a dot indicates a statistically significant association. The same approach can be used to infer associations between SLiMs and viral phenotypes, or for correlated gain/loss events. The table shows the associations observed in the E7 and E1A proteins. **D)** Functionally coupled SLiMs show coevolution signals. Upper panel: cartoons representing the interaction of E1A with the CREB binding protein (CBP) (left) and with the Retinoblastoma (Rb) protein (right). Left: The IDMBR and TRAM-CBP SLiMs in E1A mediate binding to the NCBD and TAZ2 domains of CBP . Right: The E2F, LxCxE, CKII and Acidic Region SLiMs in E1A mediate binding to the central AB domain of Rb. The E1A protein and its SLiMs are depicted in color following the color code of Figure 1. The CBP and Rb protein domains are depicted in gray. Lower panel: Associations found between the CBP- and Rb-binding SLiMs in the E1A and E7 proteins. In the lower panel the SLiMs are abbreviated as: I = IDMBR, E = E2F, T= TRAM-CBP, L = LxCxE, C = CKII, A = Acidic. Data for Panels A, C and D was taken from [47], [49] and [52].

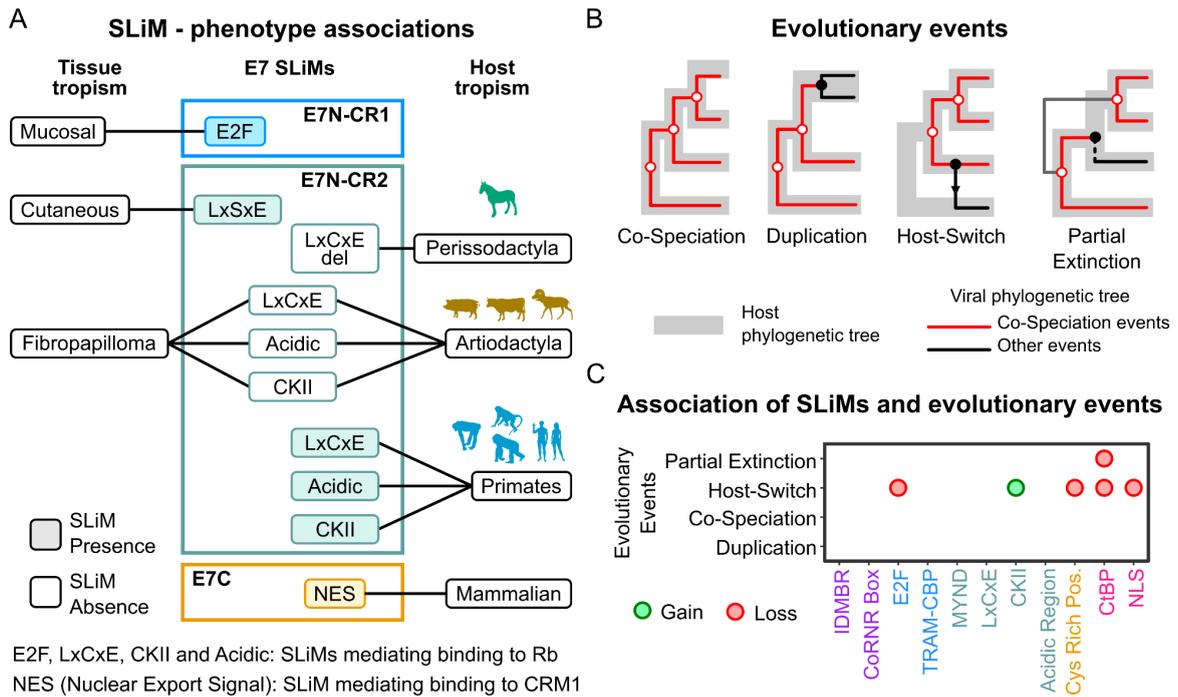

E2F, LxCxE, CKII and Acidic: SLiMs mediating binding to Rb
NES (Nuclear Export Signal): SLiM mediating binding to CRM1

**Figure 3. SLiMs are associated with viral phenotypes. A)** The presence (filled boxes) or absence (empty boxes) of five E7 SLiMs shows associations with different tissue and host tropisms. A statistically significant (p<0.05) association between a SLiM and a phenotypic trait is indicated with a line. **B)** Evolutionary events in the adenovirus phylogenetic tree indicated as thin colored lines superimposed onto the host phylogenetic tree (depicted in gray). Evolutionary events are defined in Box 1. The position of duplication, host switch and partial extinction events is indicated by a black node. **C)** The associations between E1A SLiM gain/loss events and evolutionary events are depicted as a dot-plot, where a dot indicates a statistically significant (p < 0.05) association of an evolutionary event with SLiM gain (green dot) or loss (red dot) events. The E1A SLiMs are colored following the color code of Figure 1. Data for Panel A was obtained from [47]. PanelC was adapted with permission from [49].

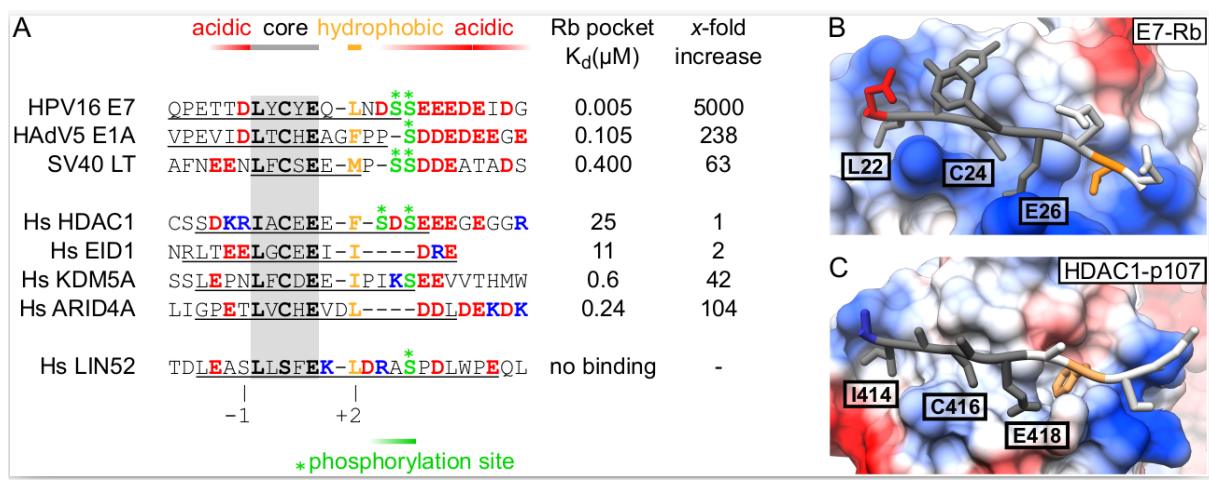

**Figure 4. SLiM core and flanking positions modulate the affinity and specificity of SLiM-mediated interactions. A)** Rb-binding affinity of LxCxE peptides from viral proteins (top) or human proteins (middle) including LIN52, a protein that preferentially binds the p107 pocket protein (bottom). X-fold increase is relative to the Hs HDAC1 protein. The sequence alignment presents the SLiM core positions in gray, with the conserved positions marked in bold text. Acidic residues in positions (-3) to (-1) before the SLiM, or after the SLiM, are colored in red and basic residues are depicted in blue. The conserved hydrophobic residue at (+2) or (+3) is colored in orange. Serines are colored green and marked with an asterisk if part of a phosphorylation site. **B-C)** Interaction between the LxCxE peptide from HPV16 E7 and human Rb (PDB: 1GUX) (B) and between the LxCxE peptide from human HDAC1 and human p107 (PDB: 7SME) (C). The core LxCxE residues are colored in gray, while flanking acidic, basic and hydrophobic residues are colored in red, blue and orange, respectively. The pocket protein surface is color coded according to the electrostatic potential of residues (acidic: red; basic: blue; neutral: white). Sidechains of R413, E417 and E419 of HDAC1 are partially missing in the structure, probably due to high flexibility. Affinity data for panel A was obtained from [59,70] (viral SLiMs) and [69] (cellular SLiMs).

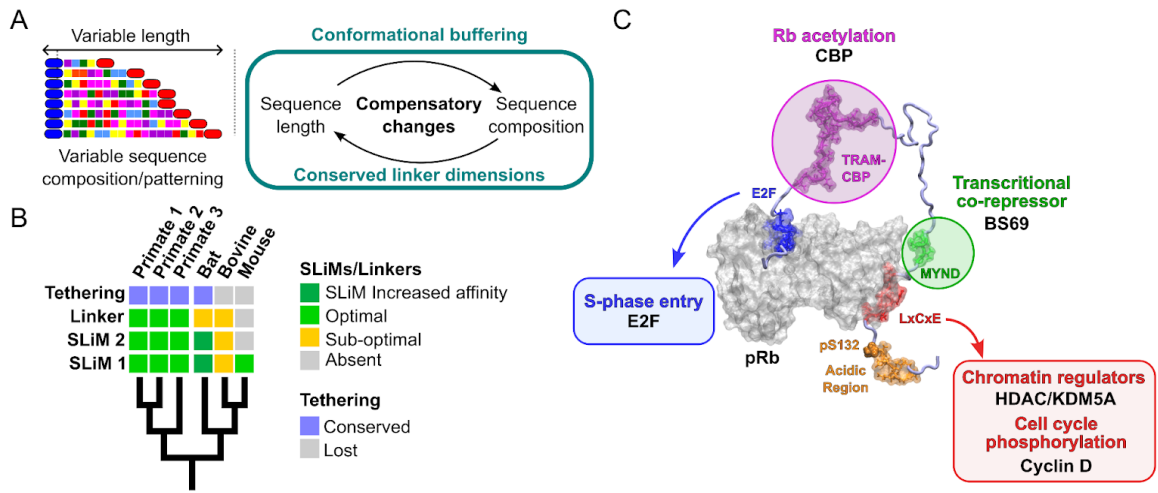

**Figure 5. Evolution of a tethering function in the adenovirus E1A linker. A)** Conserved linker dimensions are achieved by compensatory changes in linker length and sequence composition, an adaptive mechanism termed conformational buffering. **B)** Tethering is conserved through the evolution of mammalian-infecting E1A proteins. Coevolution of the linker and SLiMs is also depicted. E1A proteins infecting bats have high binding affinity SLiMs that compensate for a suboptimal linker, conserving tethering. Tethering is lost in a divergent branch that infects cows or rodents due to the presence of short linkers coupled to low affinity or missing SLiMs. **C)** Most E1A proteins conserve core functions required for cell cycle hijack (Rb binding and displacement of E2F factors and chromatin regulators). The linker contains the TRAM-CBP and MYND SLiMs that mediate binding to other targets (CBP and BS69 proteins). Sequence changes that cause the gain or loss of SLiMs in the linker of different E1A proteins could allow the E1A protein to adapt its functionality to new hosts. Panel C is adapted with permission from [59].

**Box 1. Evolution of Linear Motifs**

**Viral adaptive evolution:** It is a process through which specific phenotypes that increase fitness become fixed and maintained in a viral population.

***de novo* SLiM appearance and convergent evolution:** A SLiM is typically composed of a linear stretch of 5-10 amino acids with 3-5 conserved amino acid positions. Due to the small number of amino acids that encode a SLiM, SLiMs can appear "de novo" through one or a few point mutations, a process that is likely to occur by chance. These SLiMs can be later fixed in the population by the process of natural selection. The *de novo* appearance of a SLiM in related or unrelated proteins is an example of convergent evolution.

**Positive selection:** Positive selection is an evolutionary process characterized by high rates of nonsynonymous substitutions that increases the prevalence of traits that provide a selective advantage or improve the fitness of a population. This process is linked to adaptive evolution.

**Viral tropism:** Viral tropism can be defined as the ability of a virus to infect a specific cell type (cell tropism, i.e. T cells), a specific tissue (tissue tropism, i.e. mucosal epithelium) or a specific host species (host tropism, i.e. human). Cellular tropism, tissue tropism and host tropism can all be considered as different kinds of viral phenotypes.

**Viral evolutionary events:** Virus evolution can be studied using phylogenetic tools that allow to trace the evolutionary relationships among virus and host from the observed (or extant) sequences. This includes four events described below.

- **Cospeciation:** It is the simplest model for the coevolution of viruses and their hosts, which leads to a viral phylogeny that mirrors the host phylogeny.

- **Duplication:** A duplication occurs when the virus diverges but there is no observed divergence of the host. As a result, a new viral variant appears that continues infecting the same host.

- **Host switch:** A host switch event occurs when the virus diverges and colonizes a new host that is not closely related to the ancestral one. A host switch is also referred to as a zoonotic event or an interspecies viral transmission event.

- **Partial extinction:** A partial extinction event occurs when the virus and the host follow a pattern of cospeciation but one of the viral branches becomes extinct.

***In vitro* evolution**: *In vitro* evolution is an experimental method that allows the generation and screening of large numbers of mutants in a viral population in order to identify mutations that increase fitness or provide resistance to drugs or antibiotics.

**Author Contributions**

L.B.C. planned and supervised the work. J.G., N.P. and L.B.C. wrote and edited the manuscript. J.G. and N.P. made figures. L.B.C. and N.P. contributed to funds acquisition.

**Acknowledgements**

L.B.C. and N.P. are National Research Council Investigators (CONICET, Argentina) and J.G. is a CONICET postdoctoral fellow. This research has received funding from the European Union's Horizon 2020 research and innovation program to L.B.C. and J.G. under the Marie Skłodowska-Curie grant agreement no. 778247 (IDPfun). The work was supported by funding from Agencia Nacional de Promoción Científica y Tecnológica (ANPCyT) Grant #PICT-2017/1924 and #PICT-2019/02119 to L.B.C and #PICT-2020-SERIEA-00192 to N.P.